\documentclass[aip,jcp, reprint]{revtex4-1}
\usepackage{graphicx}
\usepackage{amsmath}
\usepackage{fancyhdr}
\usepackage{cancel}
\usepackage{color}
\usepackage{ulem}
\usepackage[breaklinks]{hyperref}
\usepackage{natbib}

\begin{document}
\title{Maximum caliber is a general variational principle for nonequilibrium statistical mechanics}

\author{Michael J. Hazoglou\dag}
\affiliation{Department of Physics \& Astronomy, Stony Brook University}
\author{Valentin Walther\dag}
\affiliation{Department of Physics \& Astronomy, Stony Brook University}
\author{Purushottam D. Dixit}
\affiliation{Department of Systems Biology, Columbia University}
\author{Ken A. Dill}
\affiliation{Department of Physics \& Astronomy, Stony Brook University}
\affiliation{Laufer Center for Physical and Quantitative Biology}
\affiliation{Department of Chemistry, Stony Brook University}

\begin{abstract}
There has been interest in finding a general variational principle for non-equilibrium statistical mechanics.  We give evidence that {\it Maximum Caliber} (Max Cal) is such a principle.  Max Cal, a variant of Maximum Entropy, predicts dynamical distribution functions by maximizing a path entropy subject to dynamical constraints, such as average fluxes.  We first show that Max Cal leads to standard near-equilibrium results -- including the Green-Kubo relations, Onsager's reciprocal relations of coupled flows, and Prigogine's principle of minimum entropy production -- in a way that is particularly simple.  More importantly, because Max Cal does not require any notion of `local equilibrium', or any notion of entropy dissipation, or even any restriction to material physics, it is more general than many traditional approaches.  We develop some generalizations of the Onsager and Prigogine results that apply arbitrarily far from equilibrium.  Max Cal is not limited to materials and fluids; it also applies, for example, to flows and trafficking on networks more broadly.

\end{abstract}

\maketitle
\dag Contributed equally

\section{Introduction}
\label{intro}

There has been interest in identifying a variational principle basis for non-equilibrium statistical mechanics (NESM).  On the one hand, phenomenological dynamics is well established.  It consists of: (a) phenomenological relationships, such as Ohm's Law of electrical current flow, Fick's Law of diffusion, Fourier's Law of conduction, the Newtonian Law of viscosity, combined with (b) conservation laws, such Kirchoff's current relationship, or similar relationships for flows of mass or heat.  The combinations of relationships of types (a) and (b) leads to well-known dynamical equations such as the Navier-Stokes and Burgers' equations of hydrodynamics or the diffusion and Smoluchowski equations, as elucidated in standard textbooks \cite{Bird2007}.

However, the search for a microscopic statistical basis for these relationships has been more challenging; see for example~\cite{bonetto2000fourier}.  While there have been many powerful and important methods for particular calculations, including the Langevin equation, Master equation, Fokker-Planck and Smoluchowski equations and others, nevertheless so far, there has been no foundational basis for NESM that provides the same power that the Second Law of thermodynamics provides for equilibria \cite{Grandy2008, Kon1998}.  Indeed, multiple context dependent variational principles have been proposed for NESM, such as Minimum Entropy Production \cite{Kon1998}, Maximum Entropy Production \cite{dewar2003information, dewar2005maximum}, and Minimum Energy Dissipation \cite{Ons1953}.  Nevertheless, a general variational quantity remains illusive.

A common basis in the above principles is to begin with the `state entropy', $S = - \sum_i p_i \log p_i$ where the $p_i$'s are the populations of equilibrium states $i = 1, 2, 3, \ldots$.  Then, a `local equilibrium' assumption is made that $S$, which is fundamentally only defined in the context of an extremum principle for predicting equilibrium, can also be a useful predictor near equilibrium.  Then the entropy production is defined as $\sigma = dS/dt$ to determines rates of approach to equilibrium or of dissipation in steady states.  One major drawback of the above methods is that they usually do not provide a natural quantifier for closeness to equilibrium; there is no systematic way to improve the near equilibrium predictions in terms of an expansion parameter.   Here, we describe an approach based on path entropies, not state entropies, that does not have these problems, is applicable even far from equilibrium, has expansion parameters, and happens to give very simple routes to deriving properties of NESM.

A goal of NESM has been to find a variational quantity which can be maximized while imposing suitable dynamical constraints. Such an approach goes beyond phenomenological descriptions of only average forces and flows and explains fluctuating quantities and higher moments of dynamical properties as well. In 1980, E.T. Jaynes proposed a candidate NESM variational principle called `Maximum Caliber' (Max Cal) \cite{Jaynes1985}.  We and others have explored its applicability as a general foundation for NESM~\cite{Pre2013,dixit2014inferring,Sto2008,wang2005maximum,Ge2012,Sto2008, Lee2012,dixit2015inferring}.    Here, we demonstrate that major results of NESM can be derived from assuming that Max Cal is a general foundational principle, these results are the Green-Kubo relations, Onsager reciprocal relations, and a generalized Prigogine's principle.

\section{Theory: the Maximum Caliber approach}
\label{theory}

\begin{figure}
        \includegraphics[scale=.2]{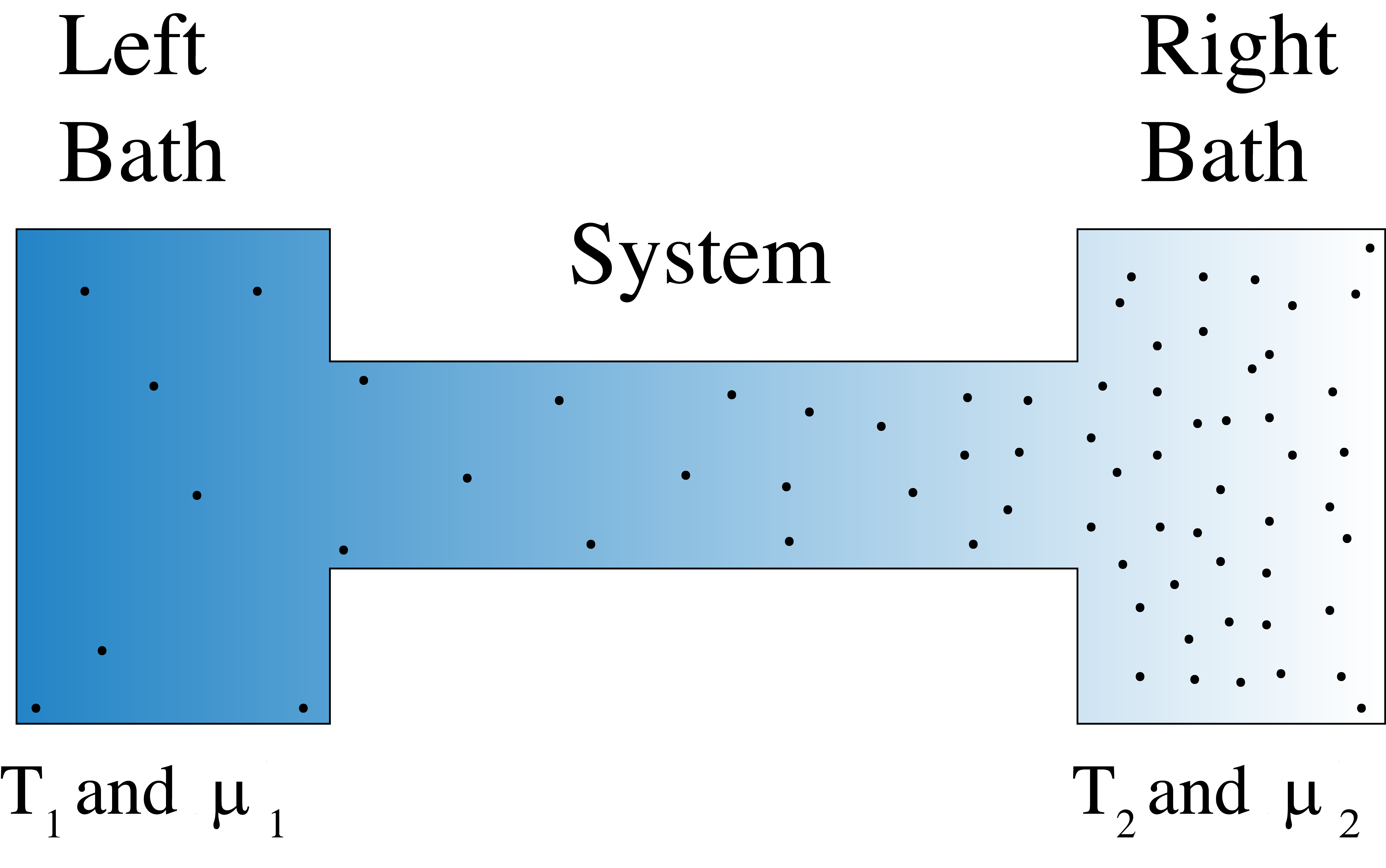}
        \caption{An illustration of coupled heat and particle flows.}
        \label{figureAA}
\end{figure}

For concreteness, we consider a discrete-time discrete-state system with two types of fluxes: of `stuff' $a$ and `stuff' $b$. Examples of $a$ and $b$ include heat, mass, electrical charge, momentum. We allow for the coupling between the two fluxes; the flux of $a$ may depend in any way on the flux of $b$. We assume that the system reaches a macroscopic nonequilibrium stationary state after a long time after it has been coupled to gradients across its boundaries that set up the fluxes.  Figure \ref{figureAA} illustrates with an example of coupled heat and particle flows in one dimension.

Suppose the system has $N$ discrete states $\{ 1, 2, \dots, N \}$. Let us define an ensemble $\{ \Gamma\} $ of  stationary-state micro-trajectories $\Gamma \equiv \dots \rightarrow i\rightarrow j \rightarrow \dots$ of fixed duration. The duration of the trajectories is not important, so we keep it unspecified. Based on how the system is coupled to the gradients and the internal structure of the system, each micro-trajectory $\Gamma$ will correspond to a specific value of flux $j_{a\Gamma}(t)$ of type $a$ (and $b$) at any time $t$. The ensemble average of the flux $J_a(t)$ of quantity $a$ at time $t$ is then given by
\begin{equation}
	J_a(t) = \langle j_{a\Gamma}(t) \rangle = \sum_\Gamma p_\Gamma j_{a\Gamma}(t) \label{eq:fluxdef}.
\end{equation}

where $p_\Gamma$ is the probability of trajectory $\Gamma$.  It is clear that different probability distributions $p_\Gamma$ could lead to macroscopic fluxes $J_a(t)$ and $J_b(t)$.  The strategy of Max Cal is to seek the particular probability distribution $p_\Gamma$ that maximizes the path entropy and is otherwise consistent with the ensemble averaged fluxes $J_a(t)$ and $J_b(t)$ for all times $t$.  We maximize the path entropy or the Caliber,
\begin{equation}
	\mathcal{C}= -\sum_\Gamma p_\Gamma \log \frac{p_{\Gamma}}{q_\Gamma}
\end{equation}
subject to normalization and flux constraints. Here, $q_\Gamma$ is the {\it reference} probability distribution of trajectories when there are no gradients, i.e. at thermodynamic equilibrium. From Fig.~\ref{figureAA}, it is clear that  there are multiple choices of $q_\Gamma$, for example, it may correspond to the equilibrium distribution at $T_1$ and $\mu_1$ or to the distribution at $T_2$ and $\mu_2$.  As an aside, the conditions for equilibrium (see below) are that (a) there are no net fluxes at equilibrium i.e.
\begin{equation}
\langle j_{a\Gamma}(t) \rangle = \sum j_{a\Gamma}(t) q_\Gamma = 0 \label{eq:noflux}
\end{equation}
 and similarly for $\langle j_{b\Gamma}(t) \rangle$ and (b) the equilibrium state satisfies microscopic reversibility: the trajectory ensemble averages at equilibrium are unchanged under path reversal. Under the situation where the flux is odd under time-reversal the condition (b) yields (a).

The Caliber \cite{Jaynes1985, Pre2013} is maximized with the constraints on the macroscopic fluxes of $a$ and $b$ at time $t$ for all $t$, which are enforced by the set of Lagrange multipliers $\{ \lambda_a(t) \}$ and $\{ \lambda_b(t) \}$:

\begin{multline}
	-\sum_\Gamma  p_\Gamma \log \frac{p_\Gamma}{q_\Gamma} +\sum_t \lambda_{a}(t) \left( \sum_{\Gamma} p_\Gamma j_{a \Gamma}(t) -J_a(t) \right)   \\ + \sum_t \lambda_{b}(t) \left( \sum_{\Gamma} p_\Gamma j_{b \Gamma}(t) -J_b(t) \right) +\alpha \left( \sum_\Gamma p_\Gamma -1\right ). \label{LagrangeFunction}
\end{multline}
Maximizing the Caliber with respect to the trajectory probability $p_\Gamma$ gives
\begin{equation}
	p_\Gamma =\frac{q_\Gamma}{Z} \exp \left( \sum_t \left[ \lambda_a(t) j_{a\Gamma}(t) +\lambda_b(t) j_{b \Gamma}(t)  \right] \right)\label{eq:pgamma}
\end{equation}
with the dynamical partition function $Z$
\begin{equation}
	Z=\sum_{\Gamma}q_\Gamma \exp \left(  \sum_t \left[ \lambda_a(t) j_{a\Gamma}(t) +\lambda_b(t) j_{b \Gamma}(t) \right] \right). \label{Partition}
\end{equation}

Equations~\eqref{eq:pgamma} and~\eqref{Partition} are the expressions of the principle of Maximum Caliber for two types of flows.  Like Maximum Entropy for equilibrium statistical mechanics, Maximum Caliber for nonequilibrium computes macroscopic quantities as derivatives of a partition-function-like quantity (in this case a sum of weights over the different pathways).  For example, average flux quantities are first derivatives of the logarithm of the {\it dynamical partition function}:
\begin{eqnarray}  
	J_a(t) = \langle j_{a\Gamma}(t) \rangle= \sum_{\Gamma}p_\Gamma j_{a \Gamma}(t)=\frac{\partial \log Z}{\partial \lambda_a (t)}.
	\label{Averages}
\end{eqnarray}
Identical equations follow for $J_b(t)$. Eq.~\eqref{Averages} allow the calculation of $\lambda_a(t)$ and $\lambda_b(t)$ from the knowledge of the functional form of $J_a [\lambda_a (t),\lambda_b (t)]$ and $J_b [\lambda_a (t),\lambda_b (t)]$ as well as the constrained values.  Higher moments of the dynamical distribution function can be calculated by taking higher derivatives of $\log Z$. For example, the second order cumulants are
\begin{eqnarray}
	\langle j_{a\Gamma}(t) j_{a\Gamma}(\tau) \rangle-\langle j_{a\Gamma}(t) \rangle \langle j_{a\Gamma}(\tau) \rangle&=&\frac{\partial \langle j_{a\Gamma}(t)\rangle}{\partial \lambda_a (\tau)}\nonumber \\ &=&\frac{\partial^2 \log Z}{\partial \lambda_a(t) \partial \lambda_a(\tau)} \\
	\langle  j_{a\Gamma}(t) j_{b\Gamma}(\tau) \rangle-\langle  j_{a\Gamma}(t) \rangle \langle  j_{b\Gamma}(\tau) \rangle &=&\frac{\partial \langle j_{a\Gamma}(t)\rangle}{\partial \lambda_b (\tau)}=\frac{\partial \langle j_{b\Gamma}(\tau)\rangle}{\partial \lambda_a (t)} \nonumber \\ &=& \frac{\partial^2 \log Z}{\partial \lambda_a(t) \partial \lambda_b(\tau)}
	\label{second order}
\end{eqnarray}
Identical expressions follow for $b$.

So far, this development is general, allowing for time-dependent fluxes.  However, for our purpose below of touching base with three well-known results of NESM, we now restrict consideration to stationary flows, $\langle j_{a\Gamma}(t) \rangle= J_a $ and $\langle j_{b\Gamma}(t) \rangle= J_b $ for all times $t$.  Appendix 1 shows that it follows that the corresponding Lagrange multipliers $\lambda_{a}$, $\lambda_{b}$ are also independent of time.  Now, we show how the Green-Kubo relations, Onsager's reciprocal relations, and Prigogine's minimum entropy production theorem, are derived quite simply from Equations~\eqref{eq:pgamma} and~\eqref{Partition}.

\subsection{Deriving the Green-Kubo Relations from Max Cal}
\label{deriveGK}

 The Green-Kubo relations are well-known expressions that give the relationships between various transport coefficients, on the one hand, and time correlation functions at equilibrium, on the other.  Here, we show that they can be derived quite directly from Max Cal.  Consider a coupled flow system in the linear regime and at stationary state when the driving forces are small. The macroscopic dynamics is time invariant (i.e. steady state). Concentrating on the fluxes at some time, call it $t=0$.   Now, expand around small driving forces $\lambda \approx 0$.  So, $\langle j_{a\Gamma}(t) \rangle = J_a(0)$ at $t=0$ is expanded to first order around $\lambda_a(\tau), \lambda_b(\tau)=0$ for all $\tau$:
\begin{eqnarray}
	J_a(0) \approx \sum_{\tau} \left[ \frac{\partial \langle j_{a\Gamma}(0) \rangle}{\partial \lambda_a(\tau)}_{\lambda=0} \lambda_a (\tau)+ \frac{\partial \langle j_{a\Gamma}(0) \rangle}{\partial \lambda_b(\tau)}_{\lambda=0} \lambda_b(\tau) \right]
	\nonumber \\
	= \lambda_a \sum_{\tau}\langle j_{a\Gamma}(0) j_{a\Gamma}(\tau) \rangle_{\lambda = 0}   + \lambda_b \sum_{\tau} \langle  j_{a\Gamma}(0) j_{b\Gamma}(\tau) \rangle_{\lambda=0}.\nonumber \\
	\label{FirstOrderExp}
\end{eqnarray}

In Eq.~\eqref{FirstOrderExp}, we recognize that at $\lambda = 0$, $p_\Gamma$ is equal to the equilibrium distribution  $ q_\Gamma$ and $\langle j_{a\Gamma(t)}\rangle =  \langle j_{b\Gamma(t)}\rangle = 0$. When the system is in steady state any time dependence of the Lagrange multipliers $\lambda_a$ and $\lambda_b$ will vanish (see Appendix~\ref{TimeIndependentLambda} for the proof).

 Eq.~\eqref{FirstOrderExp} give the Green-Kubo relations \cite{Gre1952, Gre1954, Kub1957} between the linear transport coefficient and the flux autocorrelations.  So, up to a constant factor, the Lagrange multipliers can be seen as the driving forces.

\subsection{Deriving the Onsager Reciprocal Relationships from Max Cal}

When two types of flows are linearly proportional to the two corresponding forces (i.e., in nea-requilibrium situations, then we have
\begin{eqnarray}
	J_a &=& L_{aa} \lambda_a+ L_{ab} \lambda_b \\
	J_b &=& L_{ba} \lambda_a+ L_{bb} \lambda_b.
	\label{relabel}
\end{eqnarray}
 For this situation, Onsager derived the reciprocal relationship that $L_{ab} = L_{ba}$ \cite{Ons1931a, Ons1931b}.  This result can be derived straightforwardly from Maximum Caliber.  From Eq.~\eqref{second order}, and \eqref{FirstOrderExp} above, we have

\begin{eqnarray}
        L_{ab} = \sum_\tau \frac{\partial^2 \log Z}{\partial \lambda_a(0) \partial \lambda_b(\tau)}_{\lambda=0}& =&  \sum_\tau \langle j_a(0) j_b (\tau) \rangle_{\lambda=0} \nonumber  \\ &=&  \sum_\tau \langle j_a(\tau) j_b (0) \rangle_{\lambda=0} \nonumber \\ &=& L_{ba}.
        \label{MicroReversibility}
\end{eqnarray}
The last equality follows from microscopic reversibility of the equilibrium state.  We have assumed that both fluxes have the same parity under time reversal. Summing (or integrating) over $\tau$ will not affect the symmetry of the matrix $L$.

It is clear that the symmetries in transport coefficients arise from microscopic reversibility of fluxes at equilibrium. Are there other such symmetries amongst the higher-order transport coefficients for systems not near equilibrium?  While there has been a considerable effort to discover such symmetries, no clear general results have been obtained~\cite{BK1977, BK1979,Ast2008, Art2009}. Our development shows that $n^{\rm th}$ order expansion of the flux in terms of the thermodynamic gradients will involve $(n+1)^{\rm st}$ order cumulants functions among fluxes at equilibrium. These cumulants do have some symmetry properties owing to microscopic reversibility and translational invariance with respect to time. Nonetheless, in appendix~\ref{HighOrderSym} we show that there are no simple relationships between higher-order transport coefficients.

\subsection{Deriving Prigogine's Principle of Minimum Entropy Production from Max Cal}
\label{PrigoginePrinc}

Prigogine developed a variational principle for two coupled flows near equilibrium.  If one of those flows $a$ is driven by a given force, and if the other $b$ is unconstrained, then the flux of $b$ is predicted to be that which has the minimum rate of entropy production \cite{seifert2008stochastic,tome2012entropy}.   First, here is the standard development.  If the state entropy is $S$, then the rate of entropy production in a system carrying two fluxes $J_a$ and $J_b$ is given by
\begin{equation}
\sigma = \frac{dS}{dt} = J_a \lambda_a + J_b \lambda_b
\end{equation}
where $\lambda_a$ and $\lambda_b$ are driving gradients. Using the Onsager relationships near equilibrium, we have
\begin{equation}
        \sigma=L_{aa} \lambda_a^2 +2L_{ab}\lambda_a \lambda_b+ L_{bb} \lambda_b^2.
\end{equation}

Prigogine's principle then seeks the entropy production rate that is minimal with respect to variations in $\lambda_b$,
\begin{equation}
        \frac{\partial \sigma}{\partial \lambda_b}=2(L_{ab} \lambda_a+ L_{bb} \lambda_b)=2J_b=0,
\end{equation}
which, correspondingly also predicts that $J_b = 0$ \cite{Kon1998}.

Now, here instead is the same principle derived from Max Cal.  First, express the Caliber as
\begin{eqnarray}
        \mathcal{C} &=&-\sum_\Gamma p_\Gamma \ln \left( \frac{ p_\Gamma}{q_\Gamma} \right) \nonumber \\
        &=& \ln Z- \sum_{t} \left[ \lambda_a(t)  J_a(t) + \lambda_b(t) J_b(t) \right].
\end{eqnarray}

Now, maximizing the Caliber gives
\begin{equation}
        \begin{split}
                \frac{\partial \mathcal{C}}{\partial \lambda_b(\tau)}&=-\sum_{t} \left[ \lambda_a(t) \frac{\partial J_a(t)}{\partial \lambda_b(\tau)} + \lambda_b(t)\frac{\partial J_b(t)}{\partial \lambda_b(\tau)} \right] \\
&\approx -\lambda_a L_{ab}-\lambda_b L_{bb}+\mathcal{O}(\lambda^2) =-J_b =0
        \end{split}
\end{equation}
in the linear regime this is exactly the same result as the Prigogine's principle of minimum entropy production. Max Cal makes an easily falsifiable prediction beyond the linear regime. The Caliber is maximized when
\begin{eqnarray}
\sum_{t} \left[ \lambda_a(t) \frac{\partial J_a(t)}{\partial \lambda_b(\tau)} + \lambda_b(t)\frac{\partial J_b(t)}{\partial \lambda_b(\tau)} \right]=0.\label{eq:nonlinearprig}
\end{eqnarray} 
If through detailed experiments, one knows how $J_a$ and $J_b$ depend on the imposed thermodynamic gradients $\lambda_a$ and $\lambda_b$, one can find out the gradient $\lambda_b$ to which the system adjusts itself when it is not constrained by solving Eq.~\eqref{eq:nonlinearprig}.

\section{Conclusions}
The principle of Max Cal is a putative variational principle for nonequilibrium statistical mechanics.  We show here that Max Cal provides a natural and simple route to deriving several key results of NESM, including the Green-Kubo relations, Onsager Reciprocal relations and Prigogine's minimum entropy production principle.  It's principal advantages over other derivations are that it is not limited to near equilibrium, or to `local equilibrium assumptions', has a natural `order parameter' for defining a distance from equilibrium, and has a sounder basis in principle than quantities like entropy production rates.  In short, the power of the method is its focus on path entropies, not state entropies.  We explore higher-order generalizations of Onsager relationships and Prigogine's principle for situations not near equilibrium.  

\section{Acknowledgments}  We thank Kinghuk Ghosh and Steve Presse for many deeply engaging initial discussions that led to this work.  We thank NSF grant PHY-1205881 and the Laufer Center for support to KD.

%\bibliographystyle{aipnum4-1}
%\bibliography{OnsagerReciprocalRelationsAndPrigoginesPrinciples}


\newpage
\appendix
\section{Time Independence of the Lagrange Multipliers}
\label{TimeIndependentLambda}

Here, we show $\lambda_a(t)$ and $\lambda_b(t)$ are time-independent if $J_a$ and $J_b$ are time independent. It follows from the time-independence of $J_a$, and $J_b$ and equations~\eqref{Averages} 
\begin{eqnarray}
	J_a = \frac{\partial \log Z}{\partial \lambda_a(t)}=\frac{\partial \log Z}{\partial \lambda_a(\tau)} \nonumber \\
	J_b = \frac{\partial \log Z}{\partial \lambda_b(t)}=\frac{\partial \log Z}{\partial \lambda_b(\tau)}
\end{eqnarray}
which are partial differential equations (PDEs) for the partition function $Z$ for any pair of times $t$ and $\tau$. These can be simplified by use of the chain rule
\begin{eqnarray}
\frac{1}{Z} \frac{\partial Z}{\partial \lambda_a(t)} &=& \frac{1}{Z}\frac{\partial Z}{\partial \lambda_a(\tau)} \nonumber \\
\frac{1}{Z}\frac{\partial Z}{\partial \lambda_b(t)} &=& \frac{1}{Z}\frac{\partial Z}{\partial \lambda_b(\tau)}.
\end{eqnarray}

We apply the method of separation of variables to solve these PDEs. We  assume that the solution to $Z(t)$ can be expressed as a product of arbitrary functions of each independent variable, $\prod_{t} f_t(\lambda_a(t))\cdot g_t(\lambda_b(t))$, substituting this into the PDEs above gives, 
\begin{eqnarray}
	\frac{1}{f_t(\lambda_a(t))}\frac{\partial f_t(\lambda_a(t))}{\partial \lambda_a(t)}=\frac{1}{f_\tau (\lambda_a(\tau))}\frac{\partial f_\tau (\lambda_a(\tau))}{\partial \lambda_a (\tau)}=c \qquad
	\nonumber \\
	\frac{1}{g_t(\lambda_b(t))}\frac{\partial g_t(\lambda_b(t))}{\partial \lambda_b(t)}=\frac{1}{g_\tau (\lambda_b(\tau))}\frac{\partial g_\tau (\lambda_b(\tau))}{\partial \lambda_b (\tau)}=k \qquad
\end{eqnarray}
the left and the right hand sides of these PDEs are functions of different independent variables so they must be constant (we use $c$ and $k$ as the arbitrary constants) and this is true for all times $t$ and $\tau$. The solutions are $f_t(\lambda_a (t))=a \exp(c \lambda_a(t))$ and $g_t(\lambda_b(t))=b \exp(k \lambda_b(t))$ for any time $t$, and $a$ and $b$ are arbitrary coefficients. Since these are linear PDEs the general solution is a sum of the assumed form $\prod_{t} f_t(\lambda_a(t))\cdot g_t(\lambda_b(t))$. The partition function can now be expressed in terms of the general solution and equated to the form in equation~\eqref{Partition} yielding
\begin{equation}
       \begin{split}
               & \sum_i m_i \exp \left( c_i \sum_t \lambda_a(t) + k_i \sum_t \lambda_b(t) \right) = \\
               & \sum_{\Gamma} q_\Gamma \exp \left(  \sum_t \left[ \lambda_a(t) j_{a\Gamma}(t) +\lambda_b(t) j_{b \Gamma}(t) \right] \right).
       \end{split}
\end{equation}
Identifying the index $i$ with $\Gamma$ and $m_\Gamma=q_\Gamma$, then there are two solutions to this, one is a trivial solution where $j_{a \Gamma}(t)$ and $j_{b \Gamma}(t)$ are constant in time which is not physically interesting and the other where $\lambda_a(t)$ and $\lambda_b(t)$ are time-independent.

\section{ The Lack of Symmetry Relations between Higher-Order terms from Microscopic Reversibility}
\label{HighOrderSym}
Here, we explore the question of whether there are higher-order reciprocal relations distant from equilibrium that resemble Onsager's reciprocal relations for near equilibrium.  We start with the coefficients of the second order terms. For $J_a$ this will be the following terms
\begin{multline}
               \frac{\lambda_a^2}{2}  \sum_{t, \tau}  \langle j_{a \Gamma} (0)       j_{a \Gamma} (t) j_{a \Gamma} (\tau) \rangle_{\lambda=0}   \\ + \lambda_a \lambda_b \sum_{t,\tau} \langle  j_{a \Gamma} (0)  j_{a \Gamma}(t)  j_{b \Gamma}(\tau)\rangle_{\lambda=0}  \\
 +\frac{\lambda_b^2}{2}  \sum_{t,\tau}  \langle j_{a \Gamma}(0)  j_{b \Gamma} (t)  j_{b \Gamma} (\tau) \rangle_{\lambda=0} .\label{SecondOrder}
\end{multline}
A similar expression follows for $J_b$ by replacing $j_{a\Gamma}(0)$ with $j_{b\Gamma}(0)$ in the expectation values above. It will be assumed for the rest of the section that the all fluxes are odd under time reversal (as would be the case if the transported quantities are even under time reversal, such as energy or mass) the general relations will be discussed in appendix~\ref{MixedParity}. Since the coefficients in the expansion are sums over moments we can simplify the sums and obtain relations between the moments by applying microscopic reversibility. Taking one of the moments and ordering the times as $0 \leq t \leq \tau$ microscopic reversibility gives the following expression
\begin{equation}
\langle j_{l\Gamma}(0) j_{m\Gamma}(t) j_{n\Gamma}(\tau)\rangle_{\lambda=0} = -\langle j_{n\Gamma}(0) j_{m\Gamma}(\tau-t) j_{l\Gamma}(\tau)\rangle_{\lambda=0}
\label{3rdOrderCumulantReversibility}
\end{equation}
where $l, m, n$ are either $a$ or $b$. Using this relation from time reversal several important relations can be observed.  Equation~\eqref{3rdOrderCumulantReversibility} and time translation invariance can be utilized together for more relations between cumulants in the series expansion. Shifting the time back by $\tau$ on the right hand side of equation~\eqref{3rdOrderCumulantReversibility}
\begin{equation}
\langle j_{l\Gamma}(0) j_{m\Gamma}(t) j_{n\Gamma}(\tau) \rangle_{\lambda=0} = -\langle j_{n\Gamma}(-\tau) j_{m\Gamma}(-t) j_{l\Gamma}(0) \rangle_{\lambda=0}
\label{3rdOrderCumulantShift}
\end{equation}
additionally with the time ordering $t \leq 0 \leq \tau$ gives the following result from microscopic reversibility
\begin{equation}
\langle j_{l\Gamma}(t) j_{m\Gamma}(0) j_{n\Gamma}(\tau) \rangle_{\lambda=0} = -\langle j_{n\Gamma}(-\tau) j_{m\Gamma}(0) j_{l\Gamma}(-t) \rangle_{\lambda=0}
\label{3rdOrderCumulantsAroundZero}
\end{equation}
we see by using the equations~\eqref{3rdOrderCumulantShift} and \eqref{3rdOrderCumulantsAroundZero} in equation~\eqref{SecondOrder} each term will cancel with another or vanish. So the second-order terms in $\lambda$ when both fluxes have odd parity vanish in the expansion around equilibrium. These arguments can be extended to higher-order-even terms so the fluxes will be odd functions of driving forces $\lambda_a$ and $\lambda_b$.

The coefficients of third-order terms in the $J_a$ expansion are sums over the time indices of the fourth-order cumulants of fluxes
\begin{multline}
       \langle j_{a\Gamma}(0) j_{l\Gamma}(t) j_{m\Gamma}(\tau) j_{n\Gamma}(s)\rangle_{\lambda=0} \\ - \langle j_{a\Gamma}(0) j_{l\Gamma}(t)\rangle_{\lambda=0} \langle j_{m\Gamma}(\tau) j_{n\Gamma}(s)\rangle_{\lambda=0} \\
       -\langle j_{a\Gamma}(0) j_{m\Gamma}(\tau) \rangle_{\lambda=0} \langle j_{l\Gamma}(t) j_{n\Gamma}(s)\rangle_{\lambda=0}  \\
       -\langle j_{a\Gamma}(0) j_{n\Gamma}(s) \rangle_{\lambda=0} \langle j_{l\Gamma}(t) j_{m\Gamma}(\tau)\rangle_{\lambda=0}.
\end{multline}
Here too, replacing every instance of $j_{a\Gamma}(0)$ with $j_{a\Gamma}(0)$ gives the cumulants for the third order terms for $J_b$. The last three terms when summed over time can be expressed as 
\begin{equation}
-\sum_{P} L_{al} \sum_{t,\tau}\langle j_{m\Gamma}(t) j_{n\Gamma}(\tau)\rangle_{\lambda=0}
\end{equation}
where the summation over $P$ is the set of cyclic permutations of the labels $l, m, n$. The first order coefficients make a reappearance at third-order, but this is the only nicety. The expressions for the third-order coefficients are quite cumbersome, there is no simple symmetry relations one can obtain for them. Though evaluating these coefficients can be simplified by using relations obtained with, time translation invariance, and microscopic reversibility.

\section{Various Cases of Mixed Time-Reversal Parity of Fluxes, and Higher-Order Symmetry Relations}

\label{MixedParity}
This appendix is a continuation of previous microscopic reversibility arguments made in appendix~\ref{HighOrderSym}, because so far only the case where the fluxes were negative under time-reversal were considered. To first order in the expansion of fluxes if $a$ and $b$ flows have opposite parity under time reversal we obtain by microscopic reversibility and time translation
\begin{eqnarray}
\langle j_{a\Gamma}(0) j_{b\Gamma}(\tau) \rangle_{\lambda=0} &=& -\langle j_{b\Gamma}(0) j_{a\Gamma}(\tau) \rangle_{\lambda=0} \nonumber \\
&=& -\langle j_{a\Gamma}(0) j_{b\Gamma}(-\tau) \rangle_{\lambda=0}
\end{eqnarray}
summing over $\tau$ gives the first order transport coefficients implying they all vanish $L_{ab}=L_{ba}=0$, so there can be no off diagonal coupling coefficient between flows of opposite parity.
 
Here we will start considering the second order terms for the cases when the parity of $j_{a\Gamma}$ and $j_{b\Gamma}$ are both even and when they are opposite to each. Lets define $\epsilon_a$ and $\epsilon_b$ as the time reversal parity of $j_{a\Gamma}$ and $j_{b\Gamma}$, respectively. The general form of equation~\eqref{3rdOrderCumulantReversibility} is written as
\begin{multline}
\langle j_{l\Gamma}(0) j_{m\Gamma}(t) j_{n\Gamma}(\tau)\rangle_{\lambda=0} = \\ \epsilon_l \epsilon_m \epsilon_n \langle j_{n\Gamma}(0) j_{m\Gamma}(\tau-t) j_{l\Gamma}(\tau)\rangle_{\lambda=0}
\end{multline}
and the time translation of the right hand side gives
\begin{multline}
\langle j_{l\Gamma}(0) j_{m\Gamma}(t) j_{n\Gamma}(\tau) \rangle_{\lambda=0} = \\ \epsilon_l \epsilon_m \epsilon_n \langle j_{n\Gamma}(-\tau) j_{m\Gamma}(-t) j_{l\Gamma}(0) \rangle_{\lambda=0}
\end{multline}
again with the time ordering $t \leq 0 \leq \tau$ gives the following more general result from microscopic reversibility
\begin{multline}
\langle j_{l\Gamma}(t) j_{m\Gamma}(0) j_{n\Gamma}(\tau) \rangle_{\lambda=0} = \\ \epsilon_l \epsilon_m \epsilon_n\langle j_{n\Gamma}(-\tau) j_{m\Gamma}(0) j_{l\Gamma}(-t) \rangle_{\lambda=0}.
\end{multline}

When $\epsilon_a=\epsilon_b=1$ the three equations above combine to show that for each term like $\langle j_{a\Gamma}(0) j_{a\Gamma}(t)  j_{b\Gamma}(\tau)\rangle_{\lambda=0}$, which is a term in the moment expansion of the coefficient for the $\lambda_a \lambda_b$ term in the second order expansion of $J_a$, there is an equal term in the moment expansion of the $\lambda_a^2$ coefficient for flux $J_b$, $\langle j_{b\Gamma}(0) j_{a\Gamma}(\tau-t) j_{a\Gamma}(\tau)\rangle_{\lambda=0}$ with $0 \leq t \leq \tau$. In the expansion of $J_a$ part of the $\lambda_a \lambda_b$ coefficient are $\langle j_{a\Gamma}(t) j_{a\Gamma}(0) j_{b\Gamma}(\tau) \rangle_{\lambda=0}$ with $t \leq 0 \leq \tau$. These terms equal $\langle j_{b\Gamma}(-\tau) j_{a\Gamma}(0) j_{a\Gamma}(-t) \rangle_{\lambda=0}$ and belong to the same sum for the $\lambda_a \lambda_b$ coefficient. Similar expressions follow when we switch the labels of $a$ and $b$. So we can write $J_a$ and $J_b$ to second order in $\lambda$ as
\begin{subequations}
\begin{equation}
       J_a = L_{aa}\lambda_a + L_{ab}\lambda_b + \frac{S_{a}}{2}\lambda_a^2+(K_a+V)\lambda_a\lambda_b + \frac{(P_a+M)}{2}\lambda_b^2
\end{equation}
\begin{equation}
       J_b = L_{ba}\lambda_a + L_{bb}\lambda_b+ \frac{(P_b+V)}{2}\lambda_a^2+(K_b+M)\lambda_a\lambda_b +\frac{S_{b}}{2} \lambda_b^2
\end{equation}
\end{subequations}
where $K_a, K_b, P_a, P_b, V$ and $M$ are partial summations which calculate the second order response coefficients, $S_a$ and $S_b$ give the second-order response if the other driving force is zero. Despite being able to simplify the summations with microscopic reversibility and time translation there is still no relationship between the coefficients themselves. 

For the situation of differing parity we will take $\epsilon_a=-1$ and $\epsilon_b=1$ without loss of generality since switching the labels of $a$ and $b$ will give the results for $\epsilon_b=-1$ and $\epsilon_a=1$. Moments which are odd in $j_{a\Gamma}$ will cancel with other moments of equal and opposite value in a manner similar to what was mentioned in appendix~\ref{HighOrderSym}, but $\langle j_{a\Gamma}(0) j_{a\Gamma}(t) j_{b\Gamma}(\tau)\rangle_{\lambda=0}$ = $\langle j_{b\Gamma}(0) j_{a\Gamma}(\tau-t) j_{a\Gamma}(\tau)\rangle_{\lambda=0}$ ($0 \leq t \leq \tau$) and $\langle j_{a\Gamma}(t) j_{a\Gamma}(0) j_{b\Gamma}(\tau) \rangle_{\lambda=0}$ = $\langle j_{b\Gamma}(-\tau) j_{a\Gamma}(0) j_{a\Gamma}(-t) \rangle_{\lambda=0}$ just as occurred above
\begin{subequations}
       \begin{equation}
       J_a = L_{aa}\lambda_a + (K_a+V)\lambda_a\lambda_b
       \end{equation}
       \begin{equation}
       J_b = L_{bb}\lambda_b+ (P_b+V)\lambda_a^2+S_{b} \lambda_b^2
       \end{equation}
\end{subequations}
this is effectively a hybridized version of previous results.

\end{document}